\documentclass[%
 reprint,
 amsmath,amssymb,
 aps,
]{revtex4-2}

\usepackage[utf8]{inputenc}
\usepackage{graphicx}
\usepackage{dcolumn}
\usepackage{bm}
\usepackage{hyperref}
\hypersetup{
    colorlinks=true,
    citecolor=magenta,
    colorlinks=red,
    }

\usepackage[dvipsnames]{xcolor}
\graphicspath{{figs/}}
\usepackage{xcolor}
\usepackage{physics}

\begin{document}


\title{Relaxation Control of Open Quantum Systems
}

\author{Nicolò Beato}
\email{nbeato@pks.mpg.de}
\affiliation{Max Planck Institute for the Physics of Complex Systems, N\"othnitzer Str.~38, 01187 Dresden, Germany}
\author{Gianluca Teza}
\email{teza@pks.mpg.de}
\affiliation{Max Planck Institute for the Physics of Complex Systems, N\"othnitzer Str.~38, 01187 Dresden, Germany}

\date{\today}

\begin{abstract}
    A fundamental problem in experiments with open quantum systems is to ensure steady-state convergence within a given operational time window.
    Here, we devise a general state preparation recipe to control relaxation timescales and achieve steady-state convergence within experimental run times.
    We do so by constructing a unitary operation that cancels multiple relaxation modes.
    We provide an example in a few-body interacting system (long-range qubit chain), taking into account limitations of experimentally accessible unitary operations in quantum simulators.
\end{abstract}

\maketitle

Preparing a system in a desired equilibrium configuration often requires exploring paths that are far from equilibrium.
In the presence of dissipation, the system evolves toward a steady state with a relaxation timescale set by its intrinsic properties, initial conditions, and environment \cite{van1992stochastic}.
In the context of quantum simulation on noisy intermediate-scale quantum devices \cite{preskill2018quantum}, one has to face the challenge of physical and experimental constraints limiting the overall time for a reliable observation of the dynamics \cite{nogrette2014singleatom,barredo2018synthetic,henriet2020quantum,scholl2021quantum,ebadi2022quantum,cong2022hardwareefficient,bluvstein2022quantum}.
To prepare the system in a desired physical state, it is therefore essential to design protocols that accelerate relaxation, ensuring convergence within the operating time window \cite{westhoff2025fast,zhan2025rapid,teza2025finitetemperature}.

Outside equilibrium, recent advances in tackling this fundamental problem set their roots in the Mpemba effect, a phenomenon originally related to the possibility of freezing water faster by starting from an initially hotter temperature \cite{mpemba1969cool,teza2025speedups}.
After the characterization of the Mpemba effect within a Markovian framework \cite{lu2017nonequilibrium,klich2019mpemba}, several strategies were developed to accelerate the relaxation timescale in dissipative systems in classical \cite{gal2020precooling,teza2022far,teza2023relaxation,teza2023eigenvalue} and quantum \cite{carollo2021exponentially,mori2023symmetrized,wang2024mpemba,nava2024mpemba,ruicheng2025accelerating} settings.
The idea behind these strategies is to reduce the overlap of the initial condition with the slowest relaxation mode of the dynamics, related to the second dominant eigenvalue of the evolution operator \cite{lu2017nonequilibrium,summer2025resource}.
This idea has been experimentally validated in single-particle systems, including colloidal setups \cite{kumar2020exponentially,kumar2022inverse,ibanez2024heating} and trapped ions quantum simulators \cite{aharony2024inverse,zhang2025observation}, leading to the observation of novel anomalous relaxation phenomena in classical \cite{lasanta2017when,pemartin2024shortcuts,teza2023eigenvalue,lapolla2020faster,bera2023effect,walker2023optimal}
and quantum systems, both open  \cite{tejero2024asymmetries,chatterjee2023quantum,moroder2024thermodynamics,nava2025pontus} and closed \cite{ares2023entanglement,rylands2024microscopic,ares2025quantum}.

In few- and many-body interacting systems, this simplified picture can easily break down, as the gap between the slowest mode and the following ones typically vanishes or is comparatively small with respect to their values \cite{westhoff2025fast,ruicheng2026initial}.
This is particularly fundamental for systems possessing multiple steady states, which can host phenomenology relevant for quantum technologies (e.g., dark states, unitarily evolving subspace, enhanced quantum transport \cite{buca2012,albert2016,manzano2018,dubois2023symmetry}) that remain inaccessible when the relaxation doesn't occur within the experimental time window.
Additionally, the complementary problem of extending the relaxation lifetime is also of fundamental relevance (e.g., to increase classification capabilities of quantum neural networks \cite{boneberg2025nonlinear}), but remains unaddressed to this day.

In this Letter, we introduce a general recipe for eliminating multiple relaxation modes in an open quantum system via a unitary operation, which enables the tuning of relaxation timescales without requiring any active control during the evolution.
The concurrent suppression of multiple modes unlocks a variety of applications, including, for instance, relaxation speedups even in the presence of dense, slow-decaying modes.
In our strategy, we first identify the relaxation modes that need to be suppressed to obtain the desired speedup in the relaxation dynamics and project the initial state $\rho_0$ outside of their associated subspace.
As this operation does not immediately yield a physical state (\textit{i.e.}, density matrix), we then reconstruct a state $\rho_\perp$ preserving the vanishing overlap with the undesired relaxation modes within desired precision. Finally, we build a unitary operation $U$ such that $\rho_\perp=U \rho_0 U^\dagger$.
We provide an example in a few-body interacting system (long-range qubit chain), where we investigate the effect of experimental limitations in the allowed unitary operations encountered in quantum simulators.

\begin{figure*}
\centering
\includegraphics[width=\textwidth]{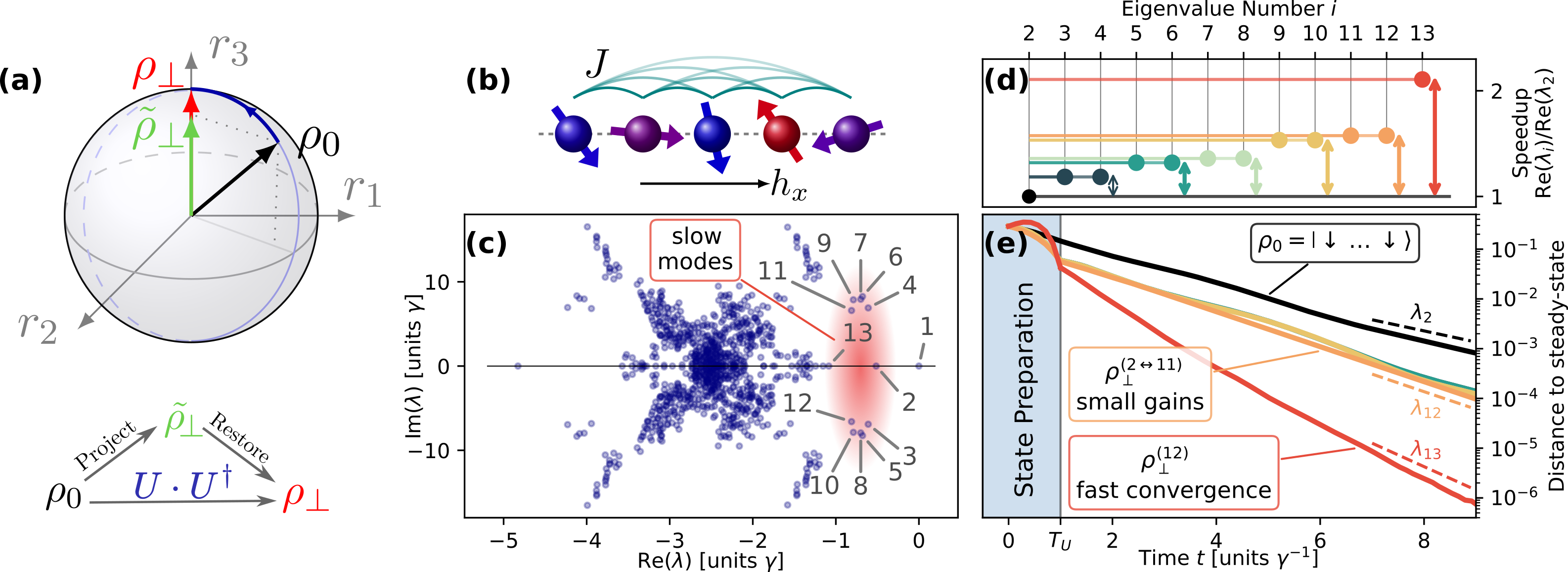}
\caption{
    Schematics representing the selective suppression of eigenmodes of the Liouvillian operator (cf.~Eq.~\eqref{eq:eigendecomp}) and exponential speedup of the relaxation.
    \textbf{(a)} A unitary rotation allows to cancel the projection of an arbitrary initial state $\rho_0$ along undesired decaying modes, which is here illustrated as the rotation of a three-dimensional vector $\rho_0$ into the vector $\rho_\perp$, perpendicular to the undesired directions $r_1,r_2$.
    \textbf{(c)} Spectrum of the Liouvillian for the long-range Ising chain sketched in panel \textbf{(b)} and defined in Eqs.~\eqref{eq:model-H} and \eqref{eq:model-L} ($N{=}5$, $h_x{=}1$, $J{=}1.25$, $\alpha{=}0.5$ and $\gamma{=}1$).
    While the relaxation timescale $\tau{=}\abs{\Re\lambda_2^{-1}}$ is regulated by the second dominant eigenvalue, eigenvalues $\lambda_{3},\dots,\lambda_{12}$ have comparable real parts (red shaded area).
    In this scenario, suppressing the second eigenmode would not provide any practical advantage.
    \textbf{(d)} Speedups (in terms of $\Re(\lambda_i)/\Re(\lambda_2)$) achievable through multi-mode suppression.
    Notice the considerable gap dividing the $12^{\textrm{th}}$ and $13^{\textrm{th}}$ eigenvalues in the example. 
    \textbf{(e)} Comparison of the different relaxations to the steady state  $\rho_{\infty}$ obtained by applying our recipe to the fully down-polarized initial state $\ket{\downarrow,\dots,\downarrow}$ (cf.~Eq.~\eqref{eq:distance_measure}).
    We selectively suppress the projection of the initial density matrix $\rho_0$ along the first $n=2,3,\dots,12$ slowest-decaying modes.
    While cancellation up to the $11^{\textrm{th}}$ mode does not sensibly change the relaxation timescale, simultaneous suppression of the first 12 modes provides a faster convergence, $\Re(\lambda_{13})/\Re(\lambda_{2}) \approx 2$.
}
\label{fig:1}
\end{figure*}

\textit{Setup---}
The dynamics of Markovian open quantum systems is described by the Gorini–Kossakowski-Sudarshan-Lindblad~(GKSL) equation \cite{gorini1976,lindblad1976,baumgartner2008,prosen2010,benatti2005,manzano2020,breuer2002theory,alicki2007,gardiner2004,rivas2012,fazio2024},
which regulates the evolution of the density matrix $\rho(t)$ in terms of the master equation $\dot \rho(t) = \mathcal L[\rho(t)]$,
\begin{equation}
    \mathcal L[\rho] = -i[H,\rho] + \sum_{i=1}^{N_J} \qty( L_i \rho L_i^\dagger - \frac12\{L_i^\dagger L_i,\rho\} ).
    \label{eq:GKSL}
\end{equation}
The Liouville superoperator $\mathcal L$  is the generator of a completely positive trace-preserving map, $H$ is the Hamiltonian of the system, and $\{L_i\}_1^{N_J}$ are $N_J\in\mathbb N$ jump operators accounting for incoherent and dissipative effects.

For a diagonalizable Liouvillian, one can find the right and left (eigen)modes $\{r_k\},\{\ell_k\}$ and eigenvalues $\{\lambda_k\}$, labeled by the index $k$ and satisfying \footnote{
Here, $\mathcal L^\dagger$ is the adjoint Lindblad map defined by $\Tr(\rho_1^\dagger \mathcal L [\rho_2])=\Tr(\mathcal L^\dagger[\rho_1^\dagger] \rho_2)$, for arbitrary density matrices $\rho_1,\rho_2$.
},
\begin{equation}
    \mathcal L [r_k] = \lambda_k r_k, \quad \mathcal L^\dagger [\ell_k] = \lambda_k^* \ell_k.
    \label{eq:eigendecomp}
\end{equation}
In general, $\lambda_k$ is complex valued and $r_k\ne\ell_k$ can be chosen to satisfy the biorthonormality relation $\Tr(\ell_k ^\dagger r_m) = \delta_{km}$ \cite{albert2018,fazio2024,kohaupt2014,supplemental}. 
In the case of real eigenvalue, $\lambda_k\in\mathbb R$, the associated modes $r_k,\ell_k$ are hermitian, $r_k=r_k^\dagger,\ell_k=\ell_k^\dagger$.
Similar to the unitary case, the nonunitary evolution operator generated by the Lindblad map can be explicitly written using its spectral decomposition as
\begin{equation}
    \exp(t \mathcal L)[\rho_0] = \sum_{k=1}^{d^2} \Tr(\ell_k^\dagger \rho_0) e^{t \lambda_k} r_k,
\end{equation}
with $d$ the Hilbert space dimension.
The complete positivity of $\mathcal L$ implies a nonpositive real part of the spectrum, allowing us to sort the eigenvalues according to $\Re(\lambda_i)\ge\Re(\lambda_{i+1})$, $i=1,\dots,d^2{-}1$. 
For bounded systems, Evans' theorem guarantees the existence of at least one steady state (or non-equilibrium steady state), associated with the vanishing eigenvalue, $\lambda_1=0$ \cite{evans1979generators,evans1977irreducible}.
Open quantum systems can generally exhibit multiple steady states \cite{manzano2018,manzano2014,buca2012,albert2014,supplemental} associated to the zero eigenvalue, and/or modes associated with purely imaginary eigenvalues, named \emph{stationary coherences} \cite{frigerio1978,vanmaassen2001,albert2018,zhang2020,thingna2021}.
In the following, we indicate with $d_s$ the number of eigenvalues with vanishing real part, and collectively refer to the associated modes as \emph{nondecaying}. 
The projection of $\rho_0$ lying on the subspace of decaying modes, $\{r_{k}\}_{k=d_s+1}^{d^2}$, is exponentially suppressed by the coefficients $\{e^{t\Re(\lambda_k)}\}_{k=d_s+1}^{d^2}$.
At sufficiently large times the slowest-decaying modes dominate the dynamics, $\rho(t) \sim \rho_{\infty}(t) + e^{-\abs{\Re(\lambda_{d_s+1})} t}r_{d_s+1}$, and the smallest decay rate $\tau^{-1} = \abs{\Re(\lambda_{d_s+1})}$ sets the system's relaxation timescale.

\textit{Speedup of relaxation---}
The basic idea behind the realization of a Mpemba effect in dissipative systems is to find an initial condition orthogonal to the slowest-decaying mode \cite{lu2017nonequilibrium,carollo2021exponentially}, which, in the case of a unique steady state, is set by the real part of the second dominant eigenvalue of the evolution operator $\lambda_2$.
However, a rather common scenario is that of a small (or missing) gap separating $\lambda_3$ from $\lambda_2$ \cite{manzano2020}, which completely undermines this strategy and its relevance for experimental contexts.
We overcome these limitations by introducing a recipe to suppress, through a unitary operation $U$, the overlap $\Tr(\ell_{a}^\dagger U \rho_0 U^\dagger)$ of the initial state $\rho_0$, with multiple modes $\ell_a {\in} \mathcal A {\subset} \{\ell_k\}_{k=1}^{d^2}$, associated with real or (pairs of) complex eigenvalues.
Theoretical restrictions and experimental challenges of our recipe are tied to physically relevant constraints and will be discussed in what follows.

Consider a finite-dimensional open quantum system, whose evolution is described by the GKSL equation [Eq.~\eqref{eq:GKSL}].
We now construct a unitary operator $U$ such that $\rho_\perp=U\rho_0U^\dagger$ satisfies $\Tr(\ell_a \rho_\perp)<\epsilon,\, \forall\,\ell_a{\in}\mathcal A$, for a positive arbitrary small $\epsilon$.
We structure the procedure leading to the construction of the operator $U$ in four main steps
\begin{center}
\includegraphics[width=.8\linewidth]{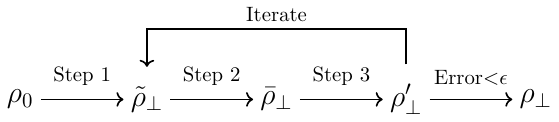}    
\end{center}
which we outline in detail below.

\noindent \textit{Step 1:~Orthogonal projection---}
First, we identify the undesired Lindblad modes $\ell_a{\in}\mathcal A$ and project the initial state $\rho_0$ outside of the associated subspace.
To this end, we expand the initial state $\rho_0$ into the Lindblad eigenbasis \footnote{
    The full diagonalization of the Liouvillian is \emph{not} required; it is sufficient to identify the (slow or fast) eigenmodes to be suppressed.
    Our discussion can be generalized to non-diagonalizable Liouvillians, mappable into Jordan canonical form; however, the recipe we present will not apply to Liouvillian modes in non-diagonalizable blocks. 
},
\begin{align}
    \rho_0 &= \sum_{k=1}^{d^2} c_k r_k, & c_k &= \Tr(\ell_k^\dagger \rho_0).
    \label{eq:rho}
\end{align}
Then, we construct the operator $\tilde\rho_\perp$
\begin{align}
    \rho_0 \mapsto \tilde\rho_\perp &= \sum_{k=1}^{d^2} \tilde c_k r_k, 
    &
    c_k \mapsto \tilde c_k &= 
    \begin{cases}
    0    & k : \ell_k \in\mathcal A \\
    c_k  & k : \ell_k \notin\mathcal A
    \end{cases},
    \label{eq:rhotilde}
\end{align}
where the coefficients associated with the undesired components are set to zero.
This procedure projects $\rho_0$ out of the subspace associated with modes $\ell_a \in \mathcal A$ but does not generally yield a density matrix.
First, $\tilde\rho_\perp$ is hermitian, $\tilde\rho_\perp = \tilde\rho_\perp^\dagger$, if and only if the chosen set $\mathcal A$ is stable under hermitian conjugation, $\mathcal A = \mathcal A^\dagger$ \footnote{
    More formally, $\ell_a^\dagger \in\mathcal A,\,\forall\, \ell_a \in\mathcal A$.
    This forbids the inclusion in $\mathcal A$ of a single element from a pair of Hermitian-conjugated left-eigenmatrices, $\{\ell_i,\ell_{i+1}\},\, \ell_{i}{=}\ell_{i+1}^\dagger$
}.
Here and in the remainder of this paper, we will make this assumption.
Second, in general, the operator $\tilde\rho_\perp$ is not a density matrix and does not possess the same spectrum as $\rho_0$; therefore, no unitary transformation $U$ satisfying $\tilde\rho_\perp = U \rho_0 U^\dagger$ exists. 
This problem is fixed in the second and third steps of the recipe, where we construct the density matrix $\rho_\perp$, which possesses the same spectrum as the initial state $
\rho_0$ and is orthogonal to the modes $\ell_a{\in}\mathcal A$ up to precision $\epsilon$.

\noindent \textit{Step 2:~Trace and purity restoration---}
We restore the trace and purity of $\rho_0$ by rescaling the coefficients of $\tilde\rho_\perp$.
We introduce two independent real parameters, $\alpha_s,\alpha \in \mathbb R$, and separately rescale the coefficients of the nondecaying and decaying modes,
\begin{align}
    \tilde\rho_\perp \mapsto \bar\rho_\perp &= \sum_{k=1}^{d^2} \bar c_k r_k, 
    &
    \tilde c_k \mapsto \bar c_k &= 
    \begin{cases}
    \alpha_s \tilde c_k & k\le d_s \\
    \alpha \tilde c_k & k>d_s
    \end{cases}.
    \label{eq:rho0bar}
\end{align}
In \cite{supplemental}, we give the expressions relating $\alpha_s,\alpha$ to the initial state and projected operator $\rho_0,\tilde\rho_\perp$, and discuss sufficient conditions for the existence of the real coefficient $\alpha$. In particular, we prove that $\alpha$ always exists in two important cases, namely when (i) only decaying modes ($\Re(\lambda_k)<0$) are suppressed or (ii) the initial state $\rho_0$ is pure.
Overall, the second step yields the operator $\bar\rho_\perp$,
orthogonal to the Liouvillian left eigenmodes $\ell_a{\in}\mathcal A$ by construction,
and satisfying the conservation of trace and purity,
\begin{align}
    1 = \Tr(\bar\rho_\perp) &= \Tr(\rho_0), & \Tr(\bar\rho_\perp^2) &= \Tr(\rho_0^2).
    \label{eq:trace-cond12}
\end{align}

\noindent \textit{Step 3: Spectrum matching---}
We enforce the equality of the spectrum between the initial and transformed states, $\rho_0,\rho_\perp$.
We diagonalize the initial density matrix $\rho_0$ and the operator $\bar\rho_\perp$ \footnote{
    In the case of a pure initial state, $\rho_0 = \dyad{\psi_0}$, it is sufficient to compute the eigenvector of $\bar\rho_\perp$ associated with the largest eigenvalue; the full diagonalization of $\bar\rho_\perp$ is not necessary. 
},
\begin{align}
    \rho_0          &= U_1 D_1 U_1^\dagger 
    &
    \bar\rho_\perp    &= U_2 D_2 U_2^\dagger,
\end{align}
and construct the density matrix
\begin{equation*}
    \rho_\perp' = U_2 D_1 U_2^\dagger.
\end{equation*}
By construction, the state $\rho_\perp'$ possesses the same spectrum as $\rho_0$. However, when $D_2 \ne D_1$, $\rho_\perp'$ is not exactly orthogonal to the set $\mathcal A$.
To this end, we therefore iterate the three steps of the recipe until numerical convergence is achieved (see \cite{supplemental} for a discussion of convergence properties). 
In this way, upon convergence of the iterative procedure within tolerance $\epsilon$, we obtain a well-defined density matrix $\rho_\perp$, which is orthogonal to the Liouvillian modes $\mathcal A$ to precision $\epsilon$ \cite{campaioli2019}. 

\noindent \textit{Step 4:~Unitary Transformation---}
Lastly, we find a unitary operator $U$ generating the transformation $\rho_\perp = U \rho_0 U^\dagger$. 
As we constructed the density matrix $\rho_\perp$ with the same spectrum as $\rho_0$, there always exists a theoretically available unitary transformation $U$ connecting the two physical states \cite{supplemental}. 
In experiments, one faces the challenge of using a limited set of generators, e.g., single-qubit operators.
This is a \emph{state-preparation problem}, to be solved using the toolbox of optimal quantum control theory \cite{koch2022,ansel2024}; the existence of a solution depends on the particular set of generators allowed by the experimental apparatus \cite{dalessandro2021,lewis2025}.
For this reason, in what follows, we restrict the discussion to a specific open quantum system and show how our recipe works in practice.

\begin{figure}
\centering
\includegraphics[width=\linewidth]{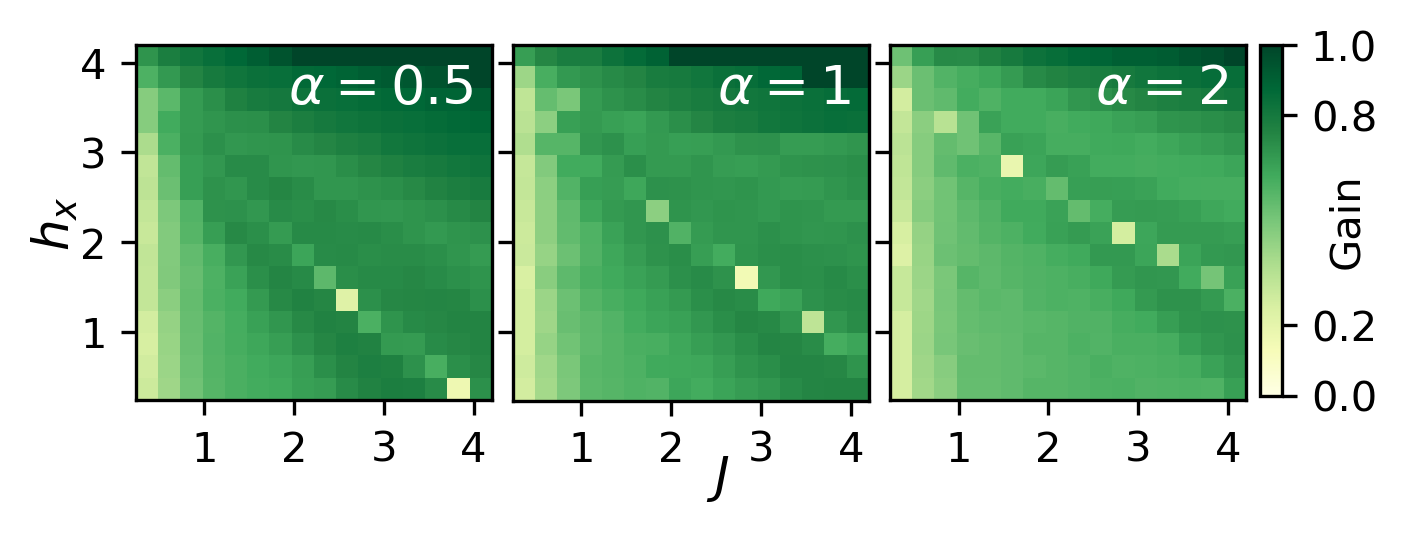}
\caption{
    Speedup of the relaxation dynamics for different parameters $\alpha,h_x,J$ of the Liouvillian in Eqs.~\eqref{eq:model-H},\eqref{eq:model-L}.
    The three figures display the relative time gain achieved by the transformed state $\rho_\perp$, as compared to the original state $\rho_0$ (cf.~Eq.~\eqref{eq:rel-time-gain}).
    The time gain is generally a significant fraction of the original time (\textit{i.e.}, 50\%), except for few isolated cases where the projection of the original state $\rho_0$ on the slowest Liouvillian eigenmodes already happens to be small.
}
\label{fig:speedup_2D_cmap}
\end{figure}

\textit{Applications---}
We demonstrate the capability of our recipe by applying it to a long-range interacting $N$-qubit chain, accessible experimentally in current trapped ions and Rydberg atom quantum simulators  \cite{nogrette2014singleatom,barredo2018synthetic,henriet2020quantum,scholl2021quantum,ebadi2022quantum,cong2022hardwareefficient,bluvstein2022quantum}. The coherent evolution is generated by the Hamiltonian
\begin{equation}
    H = h_x \sum_{i=1}^N \sigma_i^x + J \sum_{i<j}^N \frac{\sigma_i^z\ \sigma_j^z}{\abs{i-j}^\alpha}
    \label{eq:model-H}
\end{equation}
where $\sigma_i^a,a=x,y,z$ are Pauli matrices, $h_x$ sets the intensity of the external $x$-magnetic field, and $J,\alpha$ control the strength and range, respectively, of the spin-spin interaction.
The dissipative dynamics is generated by the jump operators,
\begin{align}
    L_i &= \sqrt\gamma \sigma_i^- & i\sigma_i^- &= (\sigma_i^x - \sigma_i^y)/2,
    \label{eq:model-L}
\end{align}
describing a spontaneous decay process from the state $\ket\uparrow_i$ to $\ket\downarrow_i$ with rate $\gamma$.
This model has a unique steady state, so we have $d_s=1$ (cf.~Fig.~\ref{fig:1}c).
In the following, we use the completely down-polarized initial state $\rho_0 = \ket{\downarrow,\dots,\downarrow}$.

As a first example, we engineer an exponential speedup of the relaxation toward the steady state by suppressing components of $\rho_0$ along the first slowest-decaying modes $\mathcal A_\perp^{(n)} = \{\ell_a\}_{a=2}^{n}$, for $2\le n\le12$. 
In Fig.~\ref{fig:1}c, we show the evolution of the trace norm \cite{johansson2013}
\begin{equation}    
d(\rho_1,\rho_2)= (1/2) \Tr\sqrt{\abs{\rho_1-\rho_2}^2}
\label{eq:distance_measure}
\end{equation}
of the initial and transformed states, $\rho,\rho_\perp^{(n)}$ (black and colored curves, respectively) to the steady state $\rho_\infty$.
The number of suppressed components $n$ controls the decay rate and exponentially accelerates the convergence to the steady state. 
In this example, we observe how eliminating the first ten slowest-decaying modes ($n\le11$) does not provide any substantial change in the slope of the curves of $\rho_\perp^{(n)}(t)$ (Fig. \ref{fig:1}e).
This is due to the presence of a dense cluster of slow eigenvalues $\{\lambda_k\}_{i=2}^{12}$ in the spectrum of the Liouvillian operator (Fig.~\ref{fig:1}c): only the simultaneous suppression of all such modes yields a substantial speedup.

\begin{figure}
\centering
\includegraphics[width=\linewidth]{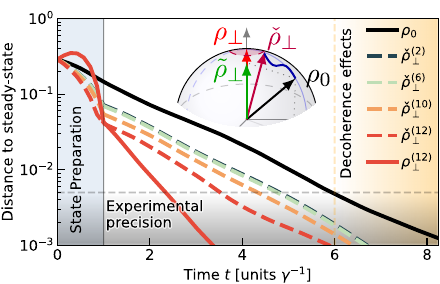}
\caption{
    Speedup of the relaxation dynamics toward the steady state, obtained through the experimentally accessible unitary transformations in Eq.~\eqref{eq:U-phys}, in the open quantum Ising chain in Eq.~\eqref{eq:model-H},\eqref{eq:model-L} ($N{=}5$, $h_x{=}1$, $J{=}1.25$, $\alpha{=}0.5$ and $\gamma{=}1$).
    The plot shows the distance to the steady state $\rho_\infty$ of the initial state $\rho_0$ and the transformed states, $\check\rho_\perp^{(n)}$, obtained with the experimentally accessible unitary (black and dashed curves, respectively; cf.~Eq.~\eqref{eq:distance_measure}).
    For reference, we also plot the evolution of $\rho_\perp^{(12)}$ from Fig.~\ref{fig:1}d, obtained with the geodesic unitary (solid red curve) \cite{supplemental}.
    Despite experimental limitations, the recipe produces a speedup in the steady-state convergence that progressively increases with the number $n$ of suppressed modes.
}
\label{fig:3}
\end{figure}

In Fig.~\ref{fig:speedup_2D_cmap}, we investigate the effectiveness of the exponential speedup recipe for different parameters of the Liouvillian. 
We vary $\alpha=0.5,1.0,2.0$, $h_x,J\in[0.25,4.0]$, and show the relative time gain achieved after suppressing up to the first 20 decaying Liouvillian modes.
In particular, we consider the times $T,T_\perp$ required by the trace norm (cf.~Eq.~\eqref{eq:distance_measure}) between $\rho_0(t),\rho_\perp(t)$ and $\rho_\infty$ to reach the experimentally motivated threshold $D_\text{min} = 10^{-3}$ and compute the relative time gain, 
\footnote{
    In Fig.~\ref{fig:3}, we neglect $T_U$ as it depends on the quantum hardware (types of gates available, duration of their execution, etc.)
    }.
\begin{equation}
    \text{Gain} = 1-T_\perp/T.
    \label{eq:rel-time-gain}
\end{equation}
The results confirm the method’s robustness and effectiveness across a wide parameter range.
Indeed, the few cases where the recipe does not achieve a considerable gain (\textit{i.e.}, $<0.1$) are due to the exceptionally small components of the original initial state $\rho_0$ onto the suppressed eigenmodes, $\ell_a \in \mathcal A$; in this rare situation, the relaxation timescale of $\rho_0$ is already fast compared to the modes we suppress.

A complementary application of our recipe is that of \textit{extending} the lifetime of the relaxation dynamics. 
In the End Matter, we show how we can suppress the projection of the initial state $\rho_0$ on all except the slowest-decaying mode(s) $\{\ell_2\}$ (or $\{\ell_2,\ell_2^*\}$ in case $\lambda_2{=}\lambda_3^*$) to increase the time needed for relaxation.
In this case, the relaxation occurring with timescale $\tau=\abs{\Re\lambda_2^{-1}}$ is not exponentially slower, but still retains a relative gain in the steady-state distance (compared to $\rho_0$) that is persistently maintained at long evolution time.

\textit{Speedup with local operations---}
In experiments, one must construct the unitary operation transforming the state $\rho_0$ into $\rho_\perp$ using only experimentally accessible operations \cite{dalessandro2021,bukov2019}.
We therefore investigate the practical utility of our recipe in trying to approximate the target states $\rho_{\perp}^{(n)}$ with the unitary operator \cite{kochsiek2022accelerating}
\begin{align}
    U(\theta,\phi) &= \bigotimes_{i=1}^N U_i(\theta,\phi)
    \label{eq:U-phys} \\
    U_i(\theta,\phi) &= \exp(i \theta \sigma_i^z / 2) \exp(i \phi \sigma_i^y / 2). \notag
\end{align}
Having only single-qubit operations, the unitary $U(\theta,\phi)$ is experimentally accessible in modern quantum devices \cite{henriet2020,tsai2022,bluvstein2022} 
\footnote{
    The choice of the \textit{Ansatz} depends on the specific problem.
    Even though more sophisticated gates can also be considered, the \textit{Ansatz} should be kept as simple as possible to minimize the preparation time, $T_U$, and simplify the optimization procedure.
}.
In particular, $U(\theta,\phi)$ uniformly rotates the qubits of the chain by the angles $\theta,\phi$ (the only two parameters of the ansatz), along the axis $\sigma_y,\sigma_z$, respectively. This operation amounts to an arbitrary reorientation of each qubit by the same angles in the Bloch sphere. 
We optimize the parameters $(\theta,\phi)$ to minimize the infidelity between $\rho_\perp$ and $\check\rho_\perp = U(\theta,\phi) \rho_0 U^\dagger(\theta,\phi)$.

In Fig.~\ref{fig:3}, we illustrate the steady-state convergence obtained preparing the system in the experimentally accessible state $\check\rho_\perp$.
Despite experimental limitations, the recipe consistently yields a speedup in the steady-state convergence that progressively increases with the number $n$ of suppressed modes.
For comparison, we also report the convergence of the state $\rho_\perp^{(12)}$ prepared with the geodesic unitary \cite{supplemental}: a mismatch in the relaxation slope is to be expected, due to the exponential precision (in evolution time $t$) required in the unitary transformation for the suppression of the coefficients $c_k$.
Nonetheless, the substantial gain in the steady-state distance (compared to $\rho_0$) provided by the preparation stage and the first phases of the relaxation is preserved throughout the entire relaxation process.
In particular, the imperfections in the preparations set a timescale $\abs{\lambda_2^{-1}}$ for all the curves, as in the ``weak'' version of the Mpemba effect \cite{klich2019mpemba,teza2025speedups}.

More importantly, quantum experiments have a finite operational time window $T_\text{max}$ and measurement sensitivity $D_\text{min}$.
As a reference, typical values for neutral-atom simulators lie on the order of $10\,\mathrm{\mu s}$ for the former and between $0.1\,\%$ and $1\,\%$ for the latter \cite{pichard2024,ebadi2022quantum,scholl2021quantum,leseleuc2018analysis}.
In Fig.~\ref{fig:3}, we highlight the experimentally inaccessible regimes with the (vertical) orange $T \gtrsim T_\text{max}$ and (horizontal) gray $D \lesssim D_\text{min}$ shaded regions; we notice that the transformed state $\check\rho_\perp^{(12)}$ reaches a threshold of distance $D_\text{min}\in[10^{-2},10^{-3}]$ significantly earlier than $\check\rho_\perp^{(2-10)}$.
This underlines how the unitary transformations, from an experimental and practical perspective, shall not necessarily reach unit fidelity in the preparation of the state $\rho_\perp$.

\textit{Discussion---}
Our method enables precise control over the relaxation dynamics of an open quantum system by selectively suppressing specific eigenmodes of the Liouvillian.
While we proved its utility focusing on the speedup and slowdown of steady-state convergence, our recipe allows targeting an \textit{arbitrary} set of eigenmodes, providing straightforward applications beyond the scenarios we presented.
Additionally, our method also works in the presence of steady-state degeneracies, making it a powerful tool for the manipulation of the dissipative dynamics in systems with nontrivial spectral structure \cite{albert2018}.
Examples include systems possessing degenerate steady states and/or a unitarily evolving subspace \cite{manzano2018,dubois2021semi,dutta2025quantum,dubois2023symmetry} and driven-dissipative systems \cite{villa2024topological,carde2025nonperturbative,banerjee2024exact}.
Although the algorithm has already delivered excellent results in the systems we analyzed (cf.~Fig.~\ref{fig:speedup_2D_cmap}), advanced update methods could improve the algorithm's convergence and accuracy \cite{quarteroni2007,broyden1965,anderson1965}.
Lastly, preservation of the initial state's spectrum corresponds to a polynomial constraint satisfaction problem \cite{byrd2003,kimura2003,chung1975,deen1971,bengtsson2011}, supporting a quest toward exact solutions \cite{gamarnik2022}.

\textit{Acknowledgments---}
The authors gratefully acknowledge the late Professor Vittorio Gorini for providing advice and useful remarks in the final stages of this project, as well as Federico Balducci, Marin Bukov, Sergio Cacciatori, Francesco Campaioli, Mattia Moroder and Ana Silva for discussions and feedback.
The authors acknowledge support from the Max Planck Society.
Numerical simulations were performed on the MPIPKS HPC cluster.

\textit{Data availability---}
The data and code associated with this manuscript are available under \cite{data_availability}.

\bibliography{refs.bib}


\onecolumngrid

\medskip


\begin{center}
    {\textbf{END MATTER: SLOWDOWN OF THE RELAXATION DYNAMICS}}
\end{center}

\medskip

\twocolumngrid

\begin{appendix}
\renewcommand{\theequation}{A\arabic{equation}}
\setcounter{equation}{0}

\label{app:relax}

\begin{figure}[h]
\centering
\includegraphics[width=.95\linewidth]{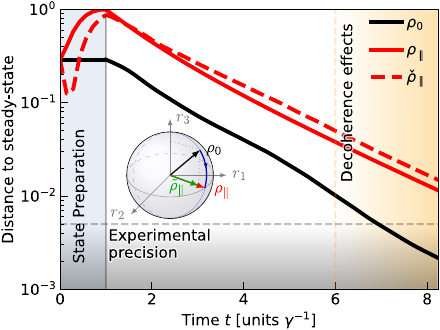}
\caption{
    Slowdown of the relaxation dynamics toward the steady state in the open quantum Ising chain in Eqs.~\eqref{eq:model-H},\eqref{eq:model-L} ($N{=}5$, $h_x{=}1$, $J{=}1.25$, $\alpha{=}0.5$ and $\gamma{=}1$).
    We use the recipe introduced in the main text to selectively suppress the projection of the initial density matrix $\rho_0$ along all except the slowest-decaying mode.
    The plot shows the distance to the steady state $\rho_\infty$ of the initial and transformed states $\rho_0(t),\rho_\parallel(t),\check\rho_\parallel(t)$ (black, solid-red and dashed-red curves, respectively).
    The dashed line refers to the state preparation obtained with the single-qubit transformation in Eq.~\eqref{eq:U-phys}.
    This result demonstrates the efficacy of the recipe for the slowdown of the relaxation dynamics.
}
\label{fig:4}
\end{figure}

Here, we demonstrate an alternative application of the recipe introduced in the main text, illustrating how the method can also be employed to slow down the relaxation dynamics. 
Here, we use the recipe to selectively suppress the projection of the initial density matrix $\rho_0$ along all except the slowest-decaying mode(s).

In Fig.~\ref{fig:4}, we demonstrate the slowdown of the relaxation dynamics toward the steady state in the open quantum Ising chain defined in Eqs.~\eqref{eq:model-H},\eqref{eq:model-L}.
This is achieved by reducing the projection of the initial state $\rho_0$ onto all except the slowest Liouvillian decaying mode(s), $\mathcal A_\parallel = \{\ell_a\}_{a>2}$ (or $\mathcal A_\parallel = \{\ell_a\}_{a>3}$, in case $\lambda_2{=}\lambda_3^*$).
The figure shows the trajectories of the original initial state $\rho_0(t)$ (black curve), the state generated by the recipe $\rho_\parallel(t)$ (solid red), and the experimentally accessible unitary $\check{\rho}_\parallel(t)$ (dashed red; cf.~Eq.~\eqref{eq:U-phys}).

To achieve a significant slowdown in the relaxation dynamics, the tolerance for the algorithm's convergence can be set smaller compared to the speedup case, e.g. $\epsilon \sim 10^{-1}$.
For a quantitative comparison, in the case of relaxation speedup, we set $\epsilon = 10^{-6}$ \cite{supplemental}.
From Fig.~\ref{fig:4}, we see that the relatively smaller tolerance allows a considerable slowdown of the relaxation process.
In particular, we observe that the distance between the transformed state $\rho_\parallel$ and the steady state $\rho_\infty$ is almost one order of magnitude larger, as compared to the initial state $\rho_0$.
Lastly, we notice that the state prepared with the experimentally accessible unitary operation exhibits a relaxation trajectory (dashed red) qualitatively close to the theoretical unitary transformation (solid red). 

These results show how the recipe introduced in the main text can also be applied to extend the relaxation dynamics of the open quantum system.

\end{appendix}

\onecolumngrid
\clearpage

\setcounter{page}{1}
\setcounter{equation}{0}
\renewcommand{\theequation}{S\arabic{equation}}

\begin{center}
    {\large \textbf{Supplemental Material for ``Relaxation control of open quantum systems''}}
\end{center}

In this supplemental material (SM), we discuss details of the calculations presented in the main text together with additional results.

\bigskip
\twocolumngrid
\bigskip

\section{The rescaling coefficients $\alpha_s,\alpha$}
\label{app:alphas}

The first step of the recipe proposed in our work projects the initial state $\rho_0$ outside the eigenspace associated with the undesired Liouvillian modes, $\ell_a{\in}\mathcal A$.
In this section, we focus on the second step, dedicated to the restoration of the trace and purity of the initial state, following the projection $\rho_0 \mapsto \tilde\rho_\perp$ (cf.~ Eqs.~\eqref{eq:trace-cond12}).
In particular, the expressions of the rescaling coefficients $\alpha_s,\alpha$ (cf.~Eq.~\eqref{eq:rho0bar}) in terms of the initial state $\rho_0$ and intermediate operator $\tilde\rho_\perp$ are explicitly given. 

The trace, $1=\Tr(\rho_0)=\Tr(\bar\rho_\perp)$, is
restored by rescaling the steady-states' coefficients $\{c_k\}_{k\le d_s}$ with the real parameter
\begin{align}
    \alpha_s &= 1/\Tr(\tilde\rho_\perp)
    \label{eq:alpha-s}
\end{align}
This simple result holds on the fact that decaying modes $\{r_k\}_{k>d_s}$ have vanishing trace, $\Tr r_k = 0,\, \forall\,k>d_s$.  

The purity, $\Tr(\rho_0^2)=\Tr(\bar\rho_\perp^2)$, is restored by rescaling the remaining coefficients $\{\tilde c_k\}_{k>d_s}$ by either one of the two roots
\begin{align}
    \alpha_\pm &= A^{-1}\qty(B\pm\sqrt{B^2-AC}),
    \label{eq:alpha-pm}
\end{align}
defined as a function of the scalar quantities
\begin{align}
    A &= \Tr(\tilde\sigma^2) \notag\\
    B &= \Re \Tr(\alpha_s\tilde\sigma_s \tilde\sigma) \notag\\
    C &= \Tr(\alpha_s^2 \tilde\sigma_s^2 {-} \rho_0^2),
    \label{eq:ABC}
\end{align}
and the operators
\begin{align}
    \tilde\sigma_s   &= \sum_{k\le d_s} \tilde c_k r_k
    &
    \tilde\sigma   &= \sum_{k>d_s} \tilde c_k r_k.
    \label{eq:sigmas}
\end{align}
satisfying $\tilde\rho_\perp = \tilde\sigma_s + \tilde\sigma$. 
By construction, if $\mathcal A = \mathcal A^\dagger$ then the operators $\tilde\sigma_s,\tilde\sigma$ (and $\tilde\rho_\perp$) are hermitian and the quantities $A,B,C \in \mathbb R$. Section \ref{app:dsc} discusses the reality condition of the parameters $\alpha_\pm$ and the geometrical interpretation associated with the rescaling operation.

\begin{figure*}
\centering
\includegraphics[width=0.9\textwidth]{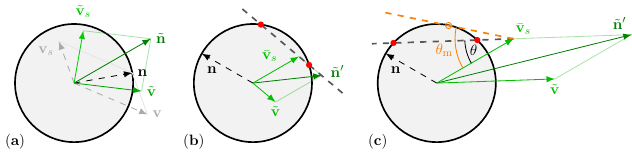}
\caption{
Schematic representation of the projection and rescaling operations for the control of the relaxation timescale (see main text).
The real vectors $\vb n,\tilde{\vb n},\dots$ represent different operators $\rho_0,\tilde\rho_\perp,\dots$ on the basis of traceless hermitian matrices $\vb S$. 
Here, we represent vectors on the two-dimensional plane $\mathcal X$ spanned by $\tilde{\vb v}_s,\tilde{\vb v}$; we use dashed arrows for vectors $\vb n, \vb v_s, \vb v$, as they do not lie on the plane $\mathcal X$.
Panel \textbf{(a)} represents the projection operation $\rho_0 \mapsto \tilde\rho_\perp$ (step one of the recipe).
Panels \textbf{(b)} and \textbf{(c)} represent the two alternative situations occurring during the rescaling operation $\tilde\rho_\perp \mapsto \bar\rho_\perp$ (step two of the recipe).
In case \textbf{(b)}, $\abs{\bar {\vb v}_s} < \abs{\vb n}$ and the coefficients $\alpha_\pm$ are guaranteed to be real.
Geometrically, the reality of coefficients $\alpha_\pm$ is tied with the existence of intersections between the sphere of radius $\abs{\vb n}$ and the dashed straight line (passing through the endpoint of $\bar {\vb v}_s$ with direction $\tilde{\vb v}'$).
In case \textbf{(c)}, $\abs{\bar {\vb v}_s} > \abs{\vb n}$ and the coefficients $\alpha_\pm$ are real if and only if the condition \eqref{eq:reality-condition} is satisfied.
The inequality \eqref{eq:reality-condition} has a simple geometrical interpretation, and it can be rewritten as $\theta < \theta_\mathrm{m}$. Here, $\theta$ is the angle between $\bar {\vb v}_s$ and $\tilde {\vb v}'$; $\theta_\mathrm{m}$ is the angle between $\bar {\vb v}_s$ and its tangent to the sphere of radius $\abs{\vb n}$.
If $\theta > \theta_\mathrm{m}$, then there is no intersection between the sphere of radius $\abs{\vb n}$ and the dashed straight line; hence, $\alpha_\pm\not\in\mathbb R$.
}
\label{fig:dsc}
\end{figure*}

\section{Conditions for the existence of real $\alpha$}
\label{app:dsc}

In this section, we discuss the reality condition of the rescaling coefficients $\alpha=\alpha_+,\alpha_-$ (see main text and Sec.~\ref {app:alphas}).

Assuming $\mathcal A$ to satisfy $\mathcal A = \mathcal A^\dagger$ (in order to have an hermitian projected state $\tilde\rho_\perp$), the reality condition of parameters $\alpha_\pm$ solely depends on the sign of the discriminant
$$\Delta = B^2-AC.$$
The condition $\Delta > 0$ has a convenient geometrical interpretation, which we now illustrate. 

First, we consider the components of the operators $\rho_0,\tilde\rho_\perp,\bar\rho_\perp$ onto the subspaces associated to the stationary and decaying Liouvillian modes,
\begin{align*}
    \rho_0 &= \sigma_s + \sigma \\
    \tilde\rho_\perp &= \tilde\sigma_s + \tilde\sigma \\
    \tilde\rho_\perp' &= \bar\sigma_s + \tilde\sigma \\
    \bar\rho_\perp &= \bar\sigma_s + \bar\sigma,
\end{align*}
where $\sigma_s$ and $\sigma$ denote the components over the stationary ($\{r_k\}_{k\le d_s}$) and decaying ($\{r_k\}_{k> d_s}$) modes, respectively.

Second, we express the density matrices $\rho_0,\tilde\rho_\perp,\tilde\rho_\perp',\bar\rho_\perp$ over a basis of hermitian traceless matrices $\{S^\alpha\}_{\alpha=2}^{d}$ satisfying $\Tr(S_i^\dagger S_j)=\delta_{ij}$. For qubit systems, we can use the basis of Pauli strings; for arbitrary systems, we can use the basis of the generalized Gell-Mann matrices.
We obtain,
\begin{align*}
    \rho_0 &= d^{-1} + (\vb v_s + \vb v) \vdot \vb S, & \vb n = \vb v_s + \vb v \\
    \tilde\rho_\perp &= d^{-1} + (\tilde {\vb v}_s + \tilde {\vb v}) \vdot \vb S,  & \tilde {\vb n} = \tilde {\vb v}_s + \tilde {\vb v}  \\
    \tilde\rho_\perp' &= d^{-1} + (\bar {\vb v}_s + \tilde {\vb v}) \vdot \vb S,  & \tilde {\vb n}' = \bar {\vb v}_s + \tilde {\vb v}  \\
    \bar\rho_\perp &= d^{-1} + (\bar {\vb v}_s + \bar {\vb v}) \vdot \vb S,  & \bar {\vb n} = \bar {\vb v}_s + \bar {\vb v} .
\end{align*}
where we defined $\vb n \cdot \vb S = \sum_{\alpha=1}^{d^2-1} n_\alpha S^\alpha$, $\vb n \in \mathbb R^{d^2-1}$.

The recipe introduced in the main text has a simple geometrical interpretation in terms of the vectors $\{\vb n,\tilde {\vb n}, \tilde {\vb n}',\bar {\vb n}\}$ and their components $\{{\vb v}_s, {\vb v}, \tilde {\vb v}_s, \tilde {\vb v}, \bar {\vb v}_s, \bar {\vb v}\}$.
In Fig.~\ref{fig:dsc} we illustrate the first and second steps of the recipe introduced in the main text within the plane $\mathcal X$ spanned by the vectors $\tilde {\vb v}_s,\tilde {\vb v}$.

In the first step, the initial state $\rho_0$ is projected outside the space spanned by the set of Liouvillian modes $\mathcal A$, $\rho_0 \mapsto \tilde\rho_\perp$.
In Fig.~\ref{fig:dsc}a, this operation is represented by the initial and projected vectors ${\vb n},\, \tilde {\vb n}$ and their components ${\vb v}_s, {\vb v}$ and $\tilde {\vb v}_s, \tilde {\vb v}$. 
As the projection do not in general preserve the purity of $\rho_0$, the norm of the vectors ${\vb n},\tilde {\vb n}$ is different and $\tilde {\vb n}$ may not lie on the sphere with radius $\abs{\vb n}$.

In the second step, the projected state  $\tilde\rho_\perp$ is transformed to restore the trace and purity of the initial state $\rho_0$. This is done in two times, first by rescaling the component of $\tilde\rho_\perp$ lying on the subspace of stationary Liouvillian modes by a factor $\alpha_s$, $\tilde \sigma_s \mapsto \alpha_s \tilde \sigma_s = \bar \sigma_s$, and then by rescaling the component of $\tilde\rho_\perp$ lying on the subspace of decaying Liouvillian modes by a factor $\alpha_\pm$, $\tilde \sigma \mapsto \alpha_\pm \tilde \sigma = \bar\sigma$. 
In Fig.~\ref{fig:dsc}b, the latter operation is represented by the dashed straight line passing through the endpoint of $\alpha_s \tilde{\vb v}_s = \bar{\vb v}_s$ with direction $\tilde{\vb v}$. Here, two different situations may arise, depending on the relative norm of $\vb n$ and $\bar{\vb v}_s$.

When $\abs{\bar {\vb v}_s} < \abs{\vb n}$, the rescaling operation is always possible as the dashed straight line is guaranteed to intersect the sphere with radius $\abs{\vb n}$ at two points (red dots of Fig.~\ref{fig:dsc}b). Algebraically, this is equivalent to the condition 
\begin{equation*}
    \abs{\bar {\vb v}_s}^2 - \abs{\vb n}^2 = \Tr(\bar \sigma_s^2) - \Tr(\rho_0^2) = C < 0.
\end{equation*}
By definition $B^2,A>0$ (cf.~Eq.~\eqref{eq:ABC}) so that $C<0$ implies that $\Delta=B^2-AC>0$ and that the rescaling coefficients $\alpha_\pm$ exist real.

When $\abs{\bar {\vb v}_s} > \abs{\vb n}$, the rescaling operation is not always possible as the dashed straight line may have empty intersection with the sphere of radius $\abs{\vb n}$.
Geometrically, the necessary and sufficient condition for the existence of intersections is $\theta < \theta_\mathrm{m}$, where $\theta$ is the angle formed by $\bar {\vb v}_s$ and  $\tilde {\vb v}$ and $\theta_\mathrm{m}$ is the angle formed by $\bar {\vb v}_s$ and its tangent to the sphere (see Fig.~\ref{fig:dsc}c). 
Algebraically, $\theta < \theta_\mathrm{m}$ is equivalent to $\Delta > 0$, as the positivity of the discriminant can be rewritten as
\begin{align*}
    \Tr(\bar \sigma_s^2) - \Tr(\rho_0^2) < \frac{(\Tr(\tilde \sigma_s \tilde \sigma))^2}{\Tr(\tilde \sigma^2)}
\end{align*}
which is equivalent to
\begin{align}
    \cos^2\theta_{\mathrm m} &= 1 -
    \frac{\abs{\vb n}^2}{\abs{\bar {\vb v}_s}^2}
    <
    \frac{
    (\bar{\vb v}_s \vdot \tilde{\vb v})^2
    }
    {
    \abs{\bar {\vb v}_s}^2 \abs{ \tilde {\vb v}}^2
    } = \cos^2 \theta.
    \label{eq:reality-condition}
\end{align}

In conclusion, let us discuss the physical implications of the reality condition $\alpha_\pm \in \mathbb R$. 
The sufficient condition $C < 0$ means that the asymptotic purity of the transformed state $\lim_{t\to\infty} \exp^{\mathcal L t} \bar\rho_\perp = \bar\sigma_s$ is smaller than the purity of the initial state, $\rho_0$. Important cases satisfying this condition are when (i) the initial state is pure, $\Tr(\rho_0^2)=1$, or (ii) the projection $\rho_0 \mapsto \tilde\rho_\perp$ does not affect the coefficients associated with the stationary modes $\{c_k\}_{k\le d_s}$ (and the initial state has a larger purity than the asymptotic steady state). 
Condition (ii) is particularly relevant as it includes all open quantum systems with a unique steady state.

\begin{figure}
\centering
\includegraphics[width=\linewidth]{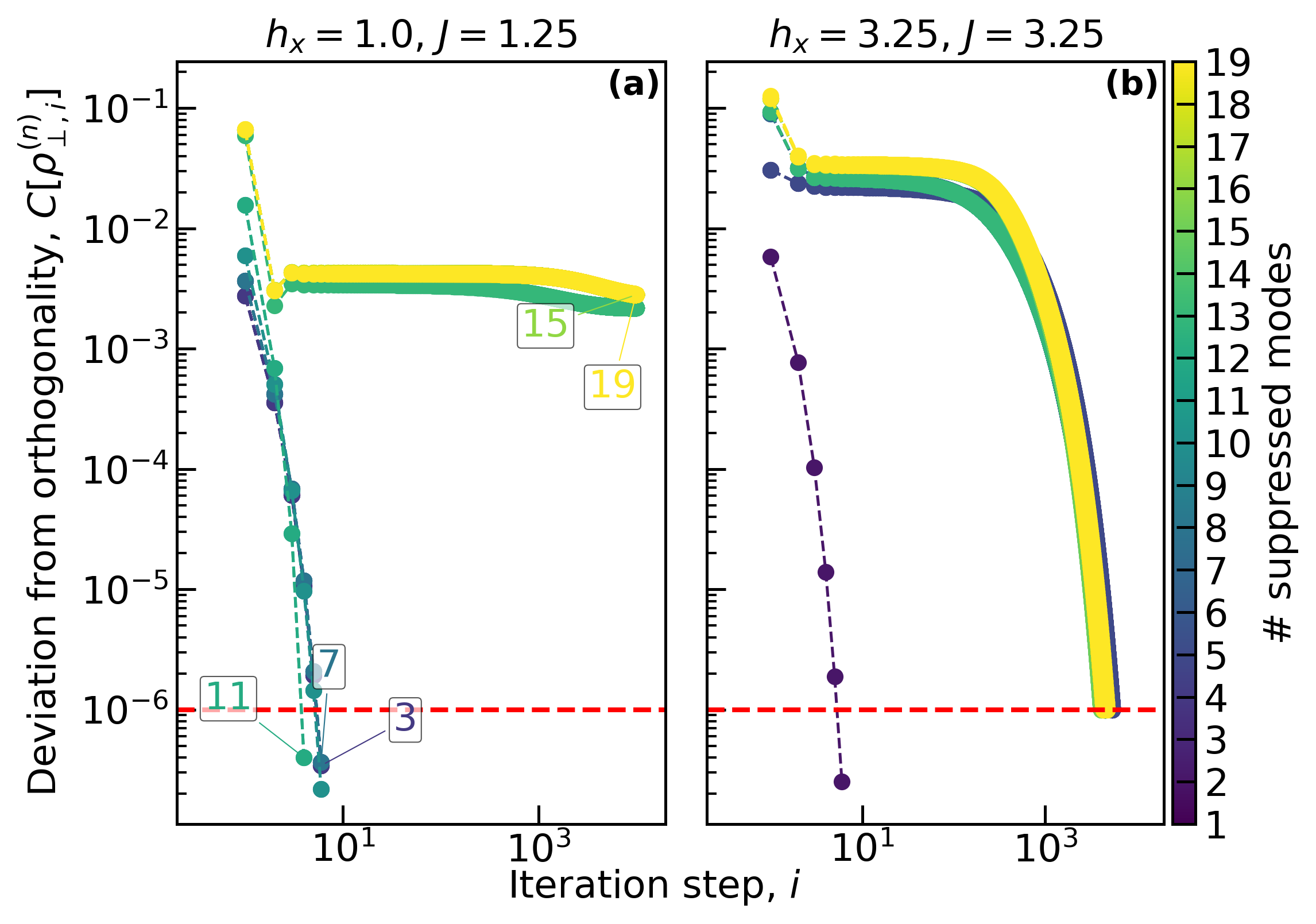}
\caption{
    Example of the evolution of the cost function $C$, defined in Eq.~\eqref{eq:cost} during the iterative procedure of the recipe introduced in the main text. 
    $C$ measures the projection of $\rho_{\perp,i}^{(n)}$ onto the Liouvillian modes $\ell_a{\in}\mathcal A$.
    The red-dashed line indicates the precision we conventionally fix, $\epsilon=10^{-6}$.
    Panel \textbf{(a)} refers to the example shown in Fig.~\ref{fig:1}.
    Observe that there exists a limited number of decaying modes that can be suppressed within the desired precision we conventionally set, $\epsilon=10^{-6}$.
    In panel \textbf{(b)}, the curves refer to the Liouvillian in Eq.~\eqref{eq:model-H},\eqref{eq:model-L} with parameters $\alpha=0.5,\, h_x=3.25,\,J=3.25$.
    From these results, we conclude that the convergence behavior of the iterative procedure depends on the details of the system considered, and that \emph{plateaus} may trap the iterations' dynamics for many ($\gtrsim10^3$) iterations.
    }
\label{fig:iteration}
\end{figure}

\begin{figure}
\centering
\includegraphics[width=\linewidth]{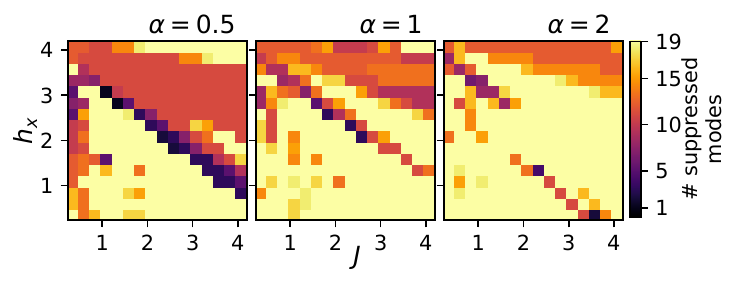}
\caption{
    Study of the convergence properties of the recipe presented in the main text, for different parameters $\alpha,h_x,J$ of the Liouvillian in Eqs.~\eqref{eq:model-H},\eqref{eq:model-L}.
    The three figures show the threshold in the number $n_*$ of slowest-decaying modes, past which the iterative procedure may fail to reach the precision we conventionally set, $\epsilon=10^{-6}$.
    The higher the threshold $n_*$, the more modes the recipe can suppress within precision $\epsilon$.
    We observe that $n_*$ largely varies for different parameters $\alpha,h_x,J$, and its range spans from a minimum of 1 to the maximum number of modes we attempted to suppress, 19.
    This result demonstrates that the number of slowest-decaying modes the recipe can cancel within a given precision is not uniform, for different Liouvillians.
    }
\label{fig:iteration-scan}
\end{figure}

\section{Convergence of the iterated procedure of the recipe}
\label{app:convergence}

In this section, we discuss the convergence properties of the iterated procedure contained in the recipe presented in the main text. 

Upon convergence of the iterative procedure, the recipe returns a well-defined density matrix $\rho_\perp$, with the same spectrum as $\rho_0$ and orthogonal to the Liouvillian modes $\ell_a \in \mathcal A$, up to the desired precision. 
We now demonstrate that the successful convergence of the iterative procedure depends on the number of decaying modes one attempts to cancel. 

At the end of each iteration, we measure the status of the iterated procedure by computing the cost function
\begin{align}
    C[\rho_\perp'] =  \sum_{\ell_a \in \mathcal A} \abs{\Tr(\ell_a^\dagger \rho_\perp')}.
    \label{eq:cost}
\end{align}
Physically, $C$ measures the projection of the state $\rho_\perp'$ onto the subspace of Liouvillian modes $\mathcal A$ we are interested in suppressing, so that it quantifies the deviation of $\rho_\perp'$ from perfect orthogonality to $\mathcal A$.

We stop the iterative procedure when one of the following three conditions is satisfied: 
(i) $C[\rho_\perp']$ decreases under the desired tolerance $\epsilon$,
(ii) the relative change of $C[\rho_\perp']$ after $\Delta I$ subsequent iterations is smaller than $p$, or 
(iii) when the maximum number of iterations $I_\text{max}$ is reached.
In our case, we fix $\epsilon=10^{-6},\, \Delta I = 10^2,\, p=10^{-6},\, I_\text{max}=10^4$.

In Fig.~\ref{fig:iteration}a, we show the evolution of the cost function $C$ during the iterative procedure for the example shown in Fig.~\ref{fig:1}.
From this result, we observe that in general there exists a threshold number $n_*$ of slowest-decaying modes that can be suppressed within the desired precision $\epsilon$. 

In Fig.~\ref{fig:iteration-scan}, we study the dependence of the threshold $n_*$ on the parameters $\alpha,h_x,J$ of the Liouvillian in Eqs.~\eqref{eq:model-H},\eqref{eq:model-L}.
We observe that $n_*$ largely varies for different parameters, and its range spans from a minimum of 1 to the maximum number of modes we attempted to suppress, 19.
This result demonstrates that the number of modes the recipe can cancel within a given precision is not uniform, for different Liouvillians.

Physically, when the recipe does not converge within precision $\epsilon$, the total deviation from orthogonality of the final state $\rho_\perp$ to the suppressed modes $\ell_a{\in}\mathcal A$ exceeds the tolerance $\epsilon$.
In this case, the suppressed modes in $\mathcal A$ contribute to the relaxation dynamics of the transformed state $\rho_\perp$ once the distance to the steady state is comparable to the precision achieved by the recipe, $d(\rho_\infty,\rho_\perp(t)) \approx \epsilon$.
From this observation, we understand that the practical usefulness of our recipe can extend beyond the cases where the iterative procedure converges to the desired precision $\epsilon$.

In Fig.~\ref{fig:iteration}b, we show the evolution of the cost function $C$ during the iterative procedure for the Liouvillian parameters $\alpha=0.5,\, h_x=3.25,\,J=3.25$.
These values are chosen to show a peculiar behavior of the cost function $C$ we observed during the iterative procedure: $C$ converges to the desired precision $\epsilon$ after an initial stabilization around a plateau, $C\approx10^{-2}$.
We observe that the convergence behavior of the iterative procedure depends on the details of the system considered. In particular, \emph{plateaus} may trap the iterations' dynamics for many ($\gtrsim10^3$) iterations.

\section{Theoretically available unitary transformations}
\label{app:theoretical}

In the main text, we give the general recipe for suppressing the components along undesired modes of the Lindblad operator in an initial state $\rho_0$.
The procedure consists of three steps and returns the state $\rho_\perp$.
In this section, we provide the expression for the theoretically available unitary transformation that returns the state $\rho_\perp$ from the initial state $\rho_0$.

Given two generic density matrices $\rho_1,\rho_2$, the necessary and sufficient condition for the existence of a unitary operation $U$ satisfying $\rho_2 = U \rho_1 U^\dagger$ is that they possess the same spectrum.

In the case of pure states, $\rho_1=\dyad{\psi_1},\rho_2=\dyad{\psi_2}$, we use the geodesic unitary $U = \exp(\theta h)$, given by
\begin{align*}
    \theta  &= \arccos{\abs{\braket{\psi_2}{\psi_1}}} \\
    h       &= \sqrt{\frac{2}{\Tr(H^\dagger H)}}H \\
    H       &= \dyad{\psi_2}{\psi_1} - \dyad{\psi_1}{\psi_2}.
\end{align*}

In the case of mixed states, a unitary transformation $U$ is constructed from the spectral decomposition of the initial and final states,
\begin{align*}
\rho_1 &= \sum_i \lambda_{1,i} \dyad{\phi_{1,i}} ,
&
\rho_2 &= \sum_i \lambda_{2,i} \dyad{\phi_{2,i}} ,
\end{align*}
as $U = \sum_i \dyad{\phi_{2,i}}{\phi_{1,i}}$.

\section{Extension to systems with multiple steady states}

In this section, we discuss in detail the extension of the recipe introduced in the main text to open quantum systems possessing multiple nondecaying states. 
In this case, the Liouvillian spectrum possesses multiple eigenvalues with zero real part, $\lambda_i {=} 0,\, i=1,\dots,d_s$. 

The eigenstates associated to these eigenvalues can be further divided into two groups: the stationary coherences and the steady states. In the first case, the imaginary part $\Im(\lambda_i)$ is strictly positive; in the second case, it is zero \cite{albert2018,albert2016,albert2014}.

The recipe presented in the main text is based on the existence of a biorthogonal set of left and right eigematrices of the Liouvillian. In general, such a set is guaranteed to exist provided the Liouvillian is diagonalizable \cite{fazio2024,albert2018}.

Unfortunately, when multiple steady states exist, standard numerical algorithms for the diagonalization of non-Hermitian matrices face important practical problems. Namely, in general (i) they do not return bi-orthogonal left- and right eigenmatrices and (ii) right eigenmatrices associated with nondecaying modes do not correspond to physical states (\textit{i.e.}, well-defined density matrices).

For this reason, in the case of degenerate steady states, the numerical diagonalization has to be followed by additional operations that solve the two aforementioned problems, and in particular:
(i) a bi-orthogonal set of left and right eigenmatrices can be extracted with a Gram-Schmidt-type algorithm \cite{kohaupt2014} or using the LU decomposition; (ii) the set of physical right eigenmatrices can be extracted either using the symmetries of the system or by first applying the Gram-Schmidt algorithm and then rotating the steady states basis to obtain positive semidefinite right-eigenmatrices \cite{thingna2021}.

\end{document}